\documentclass[twocolumn,showpacs,amssymb,prb,aps]{revtex4}
\usepackage{graphicx}
\usepackage{bm}

\begin{document}

\title{Intrinsic interface exchange coupling of ferromagnetic nanodomains in a charge ordered manganite}

\author{D. Niebieskikwiat}
\author{M.B. Salamon}
\affiliation{Department of Physics and Frederick Seitz Materials
Research Laboratory, University of Illinois at Urbana-Champaign,
Urbana, Illinois 61801, USA}

\begin{abstract}

We present a detailed magnetic study of the
Pr$_{1/3}$Ca$_{2/3}$MnO$_{3}$ manganite, where we observe the
presence of small ferromagnetic (FM) domains (diameter $\sim 10$
\AA) immersed within the charge-ordered antiferromagnetic (AFM)
host. Due to the interaction of the FM nanodroplets with a
disordered AFM shell, the low-temperature magnetization loops
present exchange bias (EB) under cooling in an applied magnetic
field. Our analysis of the cooling field dependence of the EB
yields an antiferromagnetic interface exchange coupling
comparable to the bulk exchange constant of the AFM phase. We
also observe training effect of the EB, which is successfully
described in terms of a preexisting relaxation model developed
for other classical EB systems. This work provides the first
evidence of intrinsic interface exchange coupling in phase
separated manganites.

\end{abstract}


\maketitle

\section{Introduction}

The search for improved materials for magnetic recording and
permanent magnets has been the driving force for research work in
the area of magnetic condensed matter physics.\cite{nogues99}
Artificially exchange-biased systems, like
ferromagnetic-antiferromagnetic (FM-AFM) multilayers, have been
widely studied, and they have been shown to provide large
saturation magnetizations ($M_S$) as well as high coercive fields
($H_C$).\cite{nogues99} However, the usually small exchange bias
field ($H_E$) is ultimately limited by the ability of the system
to create an uncompensated magnetization at the antiferromagnet
interface
($m_i$).\cite{nogues99,meiklejohn56,takano97,hoffmann02,antel99}
This interfacial magnetic moment induces a unidirectional
anisotropy on the ferromagnet, and depends on a number of
parameters including disorder, AFM domain- and domain-wall-size,
surface roughness, etc.\cite{malozemoff88,mauri87,keller02,liu01}
A different kind of system, known as exchange-spring magnet was
proposed by Skomski and Coey\cite{skomski93} to provide a rather
large exchange anisotropy energy. In two-phase nanocomposites,
small soft-FM inclusions adopt the anisotropy energy of a hard-FM
matrix. Such a case was observed in Nd$_2$Fe$_{14}$B/$\alpha$-Fe
alloys.\cite{lewis98} Similarly, exchange anisotropy interactions
have been studied in another large group of inhomogeneous
materials presenting FM-AFM interfaces, like
Fe/Fe-oxide,\cite{delbianco04,zheng04}
Co/CoO,\cite{meiklejohn56,skumryev03} and Ni/NiO\cite{loffler97}
nanocomposites, spin glasses, etc.\cite{nogues99}

In the heavily studied manganites, inhomogeneous magnetic phases
are known to be at the root of the colossal magnetoresistance
effect.\cite{dagotto01} Considering the presence of FM-AFM
interfaces, exchange anisotropy interactions can also be expected
in these materials. However, we are not aware of any study of
interface exchange anisotropy in inhomogeneous manganites so far.
In this paper, we present a meticulous magnetic study of the
Pr$_{1/3}$Ca$_{2/3}$MnO$_{3}$ manganite, and show clear
signatures of interface exchange coupling of FM nanodomains
immersed in the AFM background. Through an analysis of the shift
of the magnetization loops at low temperatures, we find that the
surface exchange interaction is of AFM nature, and its magnitude
is similar to the bulk exchange interaction within the AFM
volume. Finally, we also observe a training effect of the
exchange bias, which is successfully analyzed using an existing
relaxation model.

\section{Experiment}

The polycrystalline Pr$_{1/3}$Ca$_{2/3}$MnO$_{3}$ compound was
prepared by the nitrate decomposition route, as described
elsewhere,\cite{niebieskikwiat02} from high purity
Pr$_{6}$O$_{11}$, CaCO$_{3}$, and MnO. The final sintering
process was made at $1500$ $^{\circ}$C for 24 h, after which the
sample was slowly cooled to room temperature.

Magnetization ($M$) measurements were performed in a
Superconducting Quantum Interference Device (SQUID) magnetometer
with applied magnetic fields $H$ up to $70$ kOe in the
temperature range $5$ K$\leq T \leq 350$ K. The temperature
dependence of the magnetization was measured on warming with an
applied field $H=10$ kOe, in both field-cooling (FC) and
zero-field-cooling (ZFC) processes.

\section{Results and discussion}

\subsection{Observation of phase coexistence and exchange anisotropy}

Figure 1 presents data of $M/H$ as a function of temperature for
the Pr$_{1/3}$Ca$_{2/3}$MnO$_3$ manganite.  The physical
properties of this compound are mainly determined by the
electrons on the Mn ions. There is a Mn core, where three
electrons are strongly coupled to a total spin $3/2$. The
Pr$^{3+}$/Ca$^{2+}$ ratio introduces one doping electron for every
three Mn sites, responsible for the electrical transport. The
peak of $M/H$ at $T_{CO}\approx 273$ K indicates the charge
ordering transition of the material,\cite{tao05,chen97} where the
double exchange interaction is suppressed due to the localization
of the charge carriers, producing a large drop of the
susceptibility. At a lower temperature, the charge-ordered phase
undergoes a paramagnetic-to-antiferromagnetic transition at
$T_N\approx 155$ K.\cite{tao05,chen97} The inset of Fig. 1 shows
the linear $M(H)$ curve of the AFM phase at $T=5$ K. However, even
though the overall $M$-$H$ response is as expected, the sample
exhibits some peculiar behavior when is cooled with an applied
magnetic field. The curve in the inset of Fig. 1 was measured
after cooling the sample from $300$ K to $5$ K with an applied
cooling field $H_{cool}=70$ kOe. A more careful examination of
these data reveals two notable features, namely a hysteresis
(coercivity) and a shift of the $M(H)$ curve from the origin,
i.e. exchange bias. It is worth mentioning that the shift of the
loops has the same magnitude but opposite sign when the cooling
field is negative, confirming the existence of the exchange bias
phenomenon. For $H_{cool}=0$ the exchange bias effect goes away,
while the still-present coercivity is greatly reduced.

\begin{figure}
\begin{center}
\includegraphics[height=6.3cm,clip]{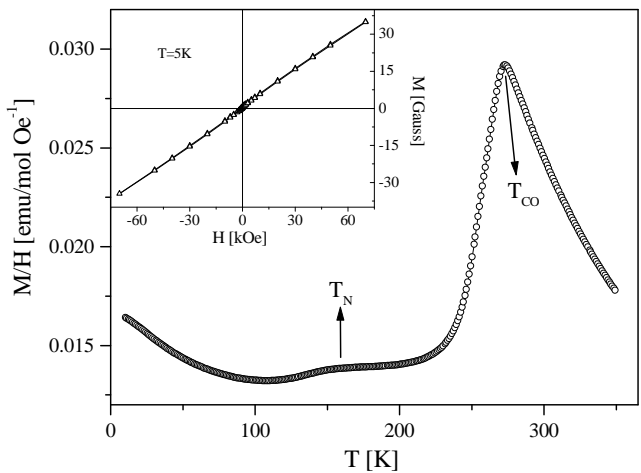}
\caption{Susceptibility ($M/H$) of the
Pr$_{1/3}$Ca$_{2/3}$MnO$_{3}$ compound as a function of
temperature ($H=10$ kOe). Inset: $M$-$H$ loop at $T=5$ K, after
field cooling the sample from $300$ K with $H_{cool}=70$ kOe.}
\label{Fig1}
\end{center}
\end{figure}

The two features clearly indicate that, superposed on the
predominant AFM signal, there is a minor FM component showing the
presence of small FM inclusions. This FM component becomes
evident in Fig. 2, where we present the resulting magnetization
($M^*$) after the AFM moment is subtracted. This intrinsic
coexistence of different magnetic phases is commonly found in
many manganese perovskite compounds in the whole range of carrier
doping, and is usually referred to as electronic phase
separation.\cite{dagotto01,niebieskikwiat02,tao05}

\begin{figure}
\begin{center}
\includegraphics[height=6.2cm,clip]{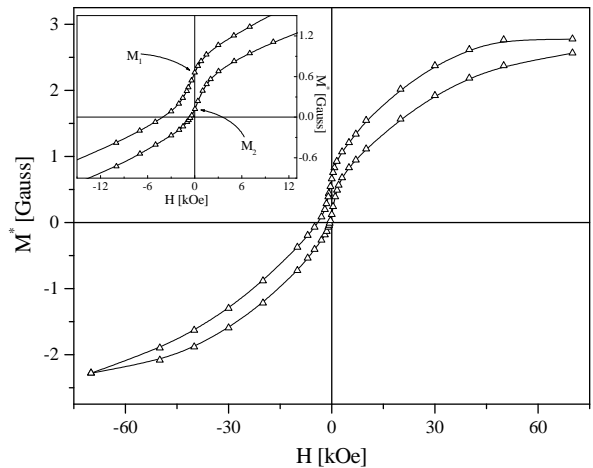}
\caption{Resulting magnetization ($M^*$) vs $H$ loop, after
subtracting the contribution of the AFM matrix ($T=5$ K,
$H_{cool}=70$ kOe). Inset: blow-up of the low-field region,
showing the determination of the remanent moments $M_1$ and
$M_2$.} \label{Fig2}
\end{center}
\end{figure}

As the magnetic field is reduced from the maximum value $70$ kOe,
the ferromagnetic moment becomes negative at $H_1\sim -3.8$ kOe,
while on the increasing branch of the loop $M^*$ changes sign at
$H_2\sim -0.5$ kOe. These two fields define a coercive field
$H_C=(H_2-H_1)/2\sim 1.6$ kOe and an exchange bias field
$H_E=(H_2+H_1)/2\sim -2.1$ kOe. However, the values of $H_1$ and
$H_2$ depend on the value of the susceptibility used to subtract
the AFM background, thus preventing a precise determination of
$H_E$. Alternatively, the exchange bias behavior can be seen as a
vertical shift of the $M^*(H)$ loop, as evidenced by the positive
remanent magnetization obtained even after recoil from a very
negative reverse field ($-70$ kOe). This effect is a
manifestation of the presence of a unidirectional exchange
anisotropy interaction, which drives the FM domains back to the
original orientation when the magnetic field is removed.
Therefore, since for $H=0$ there is no contribution from the AFM
volume, we choose to quantify the exchange bias effect using the
remanent magnetization. For this reason, we define the
magnetization shift $M_E$ and the magnetic coercivity $M_C$ as
the ``vertical axis" equivalents of $H_E$ and $H_C$,
respectively. Naming as $M_1$ and $M_2$ the remanent
magnetizations on the field decreasing and increasing branches of
the loop, respectively (as in the inset of Fig. 2), we define
$M_C=(M_1-M_2)/2$ and $M_E=(M_1+M_2)/2$ (for the data shown in
Fig. 2, $M_C\approx 0.26$ Gauss and $M_E\approx 0.40$ Gauss).
With these definitions, $M_C$ and $M_E$ are quantitative
indications of coercivity and exchange bias, respectively.
Indeed, in the simplest model of single-domain FM particles (that
we assume in this work), a relationship between $M_E$ and $H_E$
can be easily derived. Considering the magnetic moments switching
by thermal activation over the anisotropy barrier $K V$ ($K$ is
the anisotropy constant and $V$ the volume of the particles), the
exchange bias field introduces an asymmetry in the activation
energy for the backward and forward switching such that

\begin{equation}
\frac{M_E}{M_S} \sim -2 \nu_o \tau e^{-\frac{KV}{k_BT}}
\sinh\left( \frac{\mu H_E}{k_BT} \right) \label{MeVsHe}
\end{equation}
where $M_S$ is the saturation magnetization, $\nu_o\sim 10^9$
sec$^{-1}$ is the switching attempt frequency,
$\tau\sim10^2-10^3$ sec is the typical measurement time, $k_B$ is
the Boltzmann constant, and $\mu$ is the magnetic moment of the
FM particles. For a small enough magnetic energy ($\mu H_E<k_BT$),
Eq. (\ref{MeVsHe}) can be rewritten as

\begin{equation}
\frac{M_E}{M_S} \propto - H_E \label{MeVsHelinear}
\end{equation}
indicating a direct equivalence between $M_E$ and $H_E$. Our
analysis in the next subsection shows that $\mu H_E/k_BT$ is
always lower than 0.9, justifying {\it a posteriori} the use of
this relationship.

\subsection{Origin of the exchange anisotropy: temperature and cooling-field dependence}

The fact that reverse fields as high as $-70$ kOe are not able to
produce a negative $M_2$, indicates that the exchange interactions
giving rise to the described effects must be of considerable
strength. The origin of such exchange anisotropy is revealed by
the temperature dependence of the magnetization shift, as
presented in Fig. 3. In this case, the sample was cooled down from
$300$ K to $5$ K with an applied field $H_{cool}=50$ kOe, and
warmed back to the measuring temperature without removing the
applied field. At the appropriate temperature, the magnetization
loop was measured between $50$ and $-50$ kOe, from which $M_E$
and $M_C$ were calculated. It can be seen that, as the temperature
increases, the magnetization shift vanishes right at the N\'{e}el
temperature of the AFM background, $T_N\sim 155$ K, while the
coercivity remains present up to the temperature marked as $T_0$,
which is $\sim 30$ K higher than $T_N$ and indicates the onset of
the FM component.

\begin{figure}
\begin{center}
\includegraphics[height=6.2cm,clip]{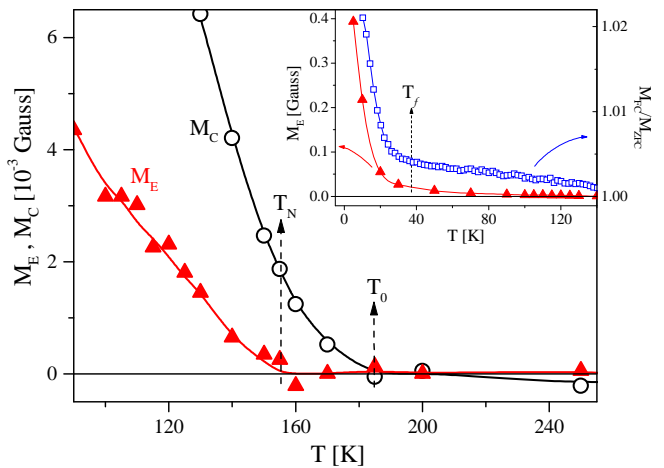}
\caption{Temperature dependence of $M_E$ and $M_C$, showing the
onset of exchange bias and coercivity at $T_N$ and $T_0$,
respectively. Inset: $M_E$ and the ratio $M_{FC}/M_{ZFC}$ at low
temperatures. Both quantities sharply increase below the freezing
temperature, $T_f\sim 35$ K.} \label{Fig3}
\end{center}
\end{figure}

The temperature evolution shown in Fig. 3 is typical of exchange
bias
systems.\cite{nogues99,meiklejohn56,takano97,hoffmann02,antel99,malozemoff88,mauri87,keller02,liu01}
As the sample is cooled through $T_N$ with an applied magnetic
field, the exchange interaction between the AFM phase and the
polarized FM phase induces an interface magnetic moment on the
antiferromagnet ($m_i$), which remains frozen at lower
temperatures.\cite{takano97,hoffmann02,nolting00} In turn, below
the freezing temperature $T_f$ ($\leq T_N$), $m_i$ provides the
necessary pinning forces (exchange anisotropy) to drive the
ferromagnet back to a positive remanence on recoil. Then, the
exchange bias behavior, observed through the appearance of a
distinctive magnetization shift, is induced by the interface
exchange coupling between the FM domains and the charge-ordered
AFM background.

Although the magnetization shift first develops at $T_N\approx
155$ K, it is useful to note that the main freezing temperature
of the system is $T_f\sim 35$ K. The inset of Fig. 3 shows the
temperature dependence of the ratio $M_{FC}/M_{ZFC}$, the
magnetization of the material measured in FC and ZFC processes,
respectively. The increase of this ratio below $T_f$ is due to
the enhancement of the magnetic irreversibility induced by the
freezing of the surface spins. Simultaneous with the upward turn
of $M_{FC}/M_{ZFC}$, $M_E$ also exhibits a steep increase of more
than one order of magnitude below $\sim 35$ K. This indicates
that a small number of interfaces start to freeze at $T_N$, but
the dominant contribution to the low temperature $M_E$ comes from
domains with $T_f\sim 35$ K. The existence of low-$T_f$ spins has
been also observed in other
systems,\cite{takano97,delbianco04,zheng04,kodama96,wang04,mitsumata03}
and have been attributed to glass-like phases occurring in
spin-disordered surfaces around the ordered
particles.\cite{delbianco04,zheng04,kodama96,wang04} Indeed, the
irreversibility shown by the ratio $M_{FC}/M_{ZFC}$ is a
fingerprint of the glassy behavior in nanoparticles systems. In
this context, a partial alignment of the glassy spins leads to the
interface moment $m_i$, and the temperature $T_f$ is a measure of
the energy barriers separating multiple spin configurations of
the glass-like phase.

In order to get some insight into the properties of this exchange
biased system, we studied the cooling field dependence of $M_E$.
For each data point shown in Fig. 4, the sample was cooled from
$300$ K to $5$ K with the applied cooling field, and a hysteresis
loop was measured between $70$ and $-70$ kOe. As shown in this
figure, $M_E$ presents a smooth increasing dependence with
$H_{cool}$, which saturates at high cooling fields.

The exchange bias field is usually thought as the balance between
the Zeeman energy of the FM particles and the surface energy due
to the interface exchange
interaction,\cite{meiklejohn56,malozemoff88} i.e.,

\begin{equation}
N_v g\mu_B H_E = - N_i J_i \frac{m_i}{g\mu_B} \label{Hevsmi}
\end{equation}

$N_v$ and $N_i$ are the number of spins inside the FM volume and
on the disordered AFM shell, respectively, thus $H_E$ depends, as
expected, on the surface-to-volume ratio. The surface exchange
constant is represented by $J_i$, $g\approx 2$ is the
gyromagnetic factor, and $\mu_B$ is the Bohr magneton. From Eq.
(\ref{Hevsmi}), the usual Meiklejohn-Bean
expression\cite{meiklejohn56} for FM-AFM bilayers is easily
recovered by making the appropriate replacements, e.g. $N_i/N_v
\rightarrow a/t_F$, where $a$ is the out-of-plane lattice
parameter and $t_F$ the thickness of the FM layer.

\begin{figure}
\begin{center}
\includegraphics[height=6.4cm,clip]{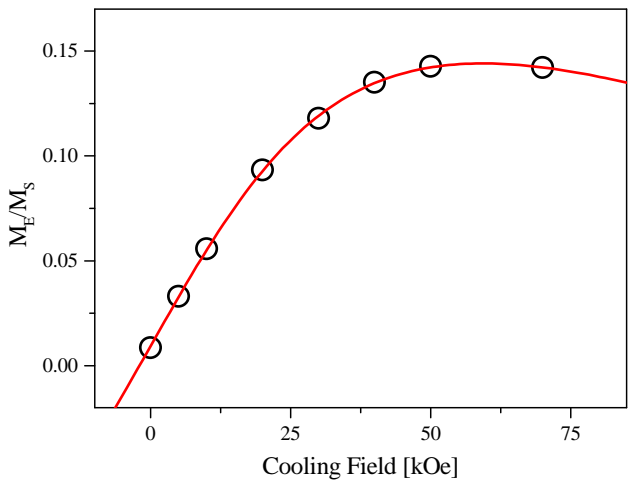}
\caption{Cooling field dependence of $M_E$. The solid line is a
fit with an exchange interaction model between the FM domains and
a disordered AFM shell (see text).} \label{Fig4}
\end{center}
\end{figure}

For small cooling fields, the coupling of the interface moment to
the FM volume dominates over the coupling of $m_i$ to $H_{cool}$,
implying that the surface spin configuration minimizes the
surface energy and $J_im_i>0$, producing a negative $H_E$.
However, it has been shown that for AFM couplings ($J_i<0$), a
large enough cooling field could prevail and flip the interface
moment, producing a positive $m_i$ at the freezing temperature
and changing the sign of the exchange bias field.\cite{nogues96}
Assuming 1-to-1 core-shell spin interactions, these competing
energies can be described in an effective magnetic field acting
on the interface moment during the freezing process as

\begin{equation}
H_{ef} = \frac{J_i}{(g\mu_B)^2}m_F(T_f) + H_{cool} \label{Hef}
\end{equation}
with $m_F(T_f)$ the magnetic moment per FM spin at the freezing
temperature. For our small magnetic domains, we assume $m_F(T_f)
= \mu_o\, L(x)$, where $\mu_o\approx 3$ $\mu_B$ is the magnetic
moment of the Mn core spin, $L(x)$ is the Langevin function,
$x=\mu H_{cool}/k_BT_f$, and $\mu=N_v\mu_o$. At lower
temperatures, because of the glassy behavior of the surface spins
the interface moment $m_i$ ($\propto H_{ef}$) remains frozen,
inducing the exchange bias response. Therefore, combining Eqs.
(\ref{MeVsHelinear}), (\ref{Hevsmi}), and (\ref{Hef}) we obtain

\begin{equation}
-H_E \propto \frac{M_E}{M_S} \propto J_i \left[
\frac{J_i\mu_o}{(g\mu_B)^2} L \left( \frac{\mu H_{cool}}{k_BT_f}
\right) + H_{cool}\right] \label{Hcooldependence}
\end{equation}

In this equation, the competition between the exchange interaction
and the cooling field becomes evident. For small $H_{cool}$ the
first term usually dominates, and $H_E$ ($<0$) depends on $J_i^2$.
However, for large cooling fields the second term ($\propto J_i$)
becomes important, and for $J_i<0$ the absolute value of $H_E$
could decrease\cite{delbianco04} or even more, it could change
sign, as observed in previous works.\cite{nogues96} In Fig. 4,
the solid line corresponds to a fit to Eq.
(\ref{Hcooldependence}) with $J_i$ and $N_v$ adjustable
parameters, which indicates a quite good agreement of the
experimental data with the simple model described above.
Moreover, the agreement is not only qualitative but also
quantitative. The exchange constant obtained from the fit was
$J_i\approx -1.7$ meV. This coupling constant indicates an AFM
coupling between the FM domains and the AFM host, explaining the
tendency of $M_E/M_S$ towards a reduction at high $H_{cool}$. On
top of this, a mean field estimation of the exchange interaction
for an antiferromagnet with $T_N=155$ K gives $J\approx -1.8$
meV, in excellent agreement with the value obtained. This
indicates that the exchange interaction within the AFM volume
extends across the boundary with the FM domains without
noticeable changes. Further, the number of spins per FM droplet
is $N_v\approx 11$ ($\mu \approx 33$ $\mu_B$), corresponding to a
droplet diameter $D\sim 10$ \AA. This kind of FM nanodomain was
already observed in the La$_{1-x}$Ca$_{x}$MnO$_3$
manganite\cite{granado03} and in
La$_{0.94}$Sr$_{0.06}$MnO$_3$,\cite{hennion00} where small angle
neutron scattering experiments show the presence of FM droplets
with the same characteristic size, and a density of domains $n\sim
10^{-5}$ \AA$^{-3}$. In our Pr$_{1/3}$Ca$_{2/3}$MnO$_3$ sample,
an estimation of the density of FM droplets can be made from the
saturation magnetization, which can be written as

\begin{equation}
M_S = n \mu \label{Ms}
\end{equation}

With $M_S\sim 2.8$ Gauss obtained from Fig. 2, we estimate $n\sim
0.9\times 10^{-5}$ \AA$^{-3}$, which is very similar to the
mentioned neutron scattering results. This density of droplets
implies that the FM phase occupies only around $0.5\%$ of the
volume of the material, explaining the almost imperceptible FM
signal that superposes on the AFM magnetization.

At this point it is worth noting that, although in previous works
the small interface moment $m_i$ has been detected by different
techniques,\cite{takano97,hoffmann02,antel99} in our case this
moment is negligible as compared to $M_E$. An estimation from Eq.
(\ref{Hevsmi}) indicates that the maximum contribution of the
interface is $m_i\approx -0.027$ $\mu_B$ per surface spin. This
represents only a $0.9\%$ of the maximum possible value ($3$
$\mu_B$), in coincidence with measurements in other different
systems.\cite{takano97} This $m_i$ is indeed much smaller than
the shift of the loops, which is $\sim 0.42$ $\mu_B$ per FM spin
($M_E/M_S\sim 0.14$) at the same cooling field. Moreover, while
$m_i$ is negative the measured $M_E$ is always positive.
Therefore, the magnetization shift measured in our system
corresponds almost in its entirety to the remanence of the FM
domains, as assumed in our analysis.

The exchange bias phenomenon observed in our material would be
similar to that taking place in artificially fabricated
nanocomposites,\cite{meiklejohn56,delbianco04,zheng04,skumryev03,loffler97}
where FM nanoparticles are surrounded by the AFM oxide. This kind
of core-shell interaction has been shown to provide quite strong
magnetic anisotropies. In our case, using $M_E/M_S\sim 0.14$ and
$H_E\sim -2$ kOe (for $H_{cool}=70$ kOe), from Eq. (\ref{MeVsHe})
we can estimate an anisotropy constant $K\sim 3\times 10^7$
erg/cm$^3$. This is a quite important magnetic anisotropy, two
orders of magnitude larger than the usual magnetocrystalline
anisotropy measured in bulk FM manganites.\cite{lofland97} It is
clear that this notable anisotropy constant is related to the
particular magnetic configuration of the
Pr$_{1/3}$Ca$_{2/3}$MnO$_3$ compound. In spring-magnet materials,
it has been shown that the exchange anisotropy increases when the
soft-FM domain size decreases, acquiring the anisotropy energy of
the hard-FM matrix.\cite{skomski93,lewis98} Similarly, the
exchange bias phenomenon is related to the surface exchange
interaction between the FM and AFM phases. Therefore, it is
widely accepted that, in nanocomposite materials, $K$ is
proportional to the surface-to-volume ratio of the FM domains. In
the case of our nanometric FM domains, where $N_v\sim 11$, the
ratio $N_i/N_v$ is close to 1, being compatible with the high
anisotropy constant estimated. Elastic effects in the
charge-ordered phase tend to produce not spherical but elongated
secondary phases,\cite{tao05,podzorov01} therefore a shape
anisotropy energy could be also contributing to the anisotropy
constant. From the structural point of view, the strength of the
exchange anisotropy is influenced by the existence of surface
strain and atomic displacement. This disorder modifies the
interface exchange constant,\cite{malozemoff88,keller02} and
could be particularly relevant in nanocomposites where the
constituents are structurally and chemically different. However,
our manganite is an {\it intrinsic nanocomposite}, where the FM
droplets appear naturally embedded in the AFM background and
where structural and chemical differences should be less
important. As a consequence, the obtained interface exchange
constant is as large as in the bulk of the material, favoring a
high exchange anisotropy as well.

\subsection{Training effect}

\begin{figure}
\begin{center}
\includegraphics[height=6.6cm,clip]{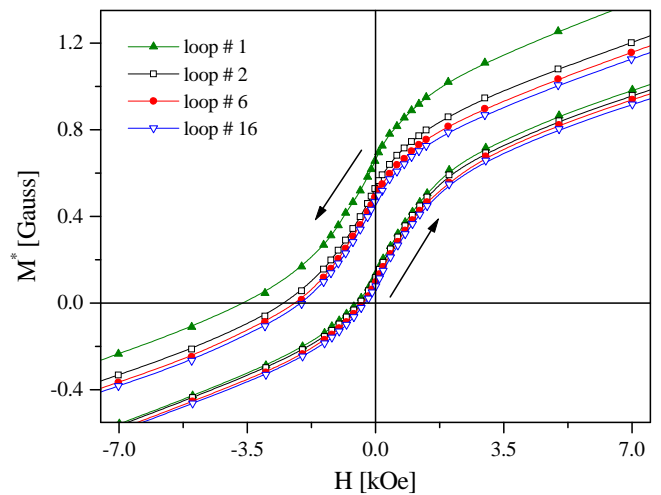}
\caption{Consecutive magnetization loops showing the training
effect of the exchange bias. These data were measured after
cooling the sample with $H_{cool}=70$ kOe.} \label{Fig5}
\end{center}
\end{figure}

In exchange bias systems, a gradual decrease of the anisotropy
interaction is commonly found as the material is continuously
field-cycled; the so-called training
effect.\cite{nogues99,keller02,zheng04,binek04,hochstrat02,hoffmann04}
As a result, the exchange bias and the coercive fields decrease
with increasing loop index number. In our manganite, the training
effect is also present, as shown in Fig. 5. This figure presents
the consecutive $M^*$-$H$ loops measured for fields between $70$
and $-70$ kOe, after the sample was field-cooled with
$H_{cool}=70$ kOe (only loops with index number $\lambda =1$,
$2$, $6$, and $16$ are shown). The relaxation of the
magnetization shift is evident and, as shown by the evolution of
$M_E$ with $\lambda$ (Fig. 6), this relaxation is particularly
important between the first and second loops, where $M_E$ falls
by $\sim 20\%$.

\begin{figure}
\begin{center}
\includegraphics[height=6.5cm,clip]{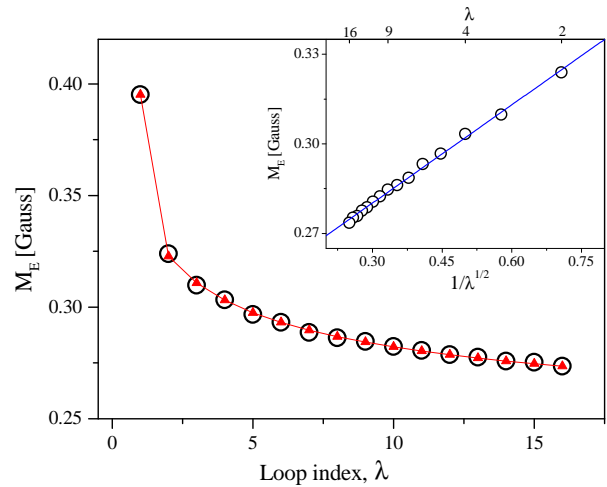}
\caption{The open circles are the experimental data of
magnetization shift ($M_E$) vs loop index number ($\lambda$). The
solid triangles correspond to a fit with a recursive law, that
results from a discretized relaxation model. Inset: $M_E$ vs
$1/\sqrt{\lambda}$, showing the linear behavior usually obtained.
The line is a least squares fit.} \label{Fig6}
\end{center}
\end{figure}

The training effect has been explained in terms of the
demagnetization of the AFM surface moment
$m_i$.\cite{zheng04,binek04,hoffmann04} As the FM domain switches
back and forth under the influence of the applied field, a
relaxation of the surface spin configuration towards the
equilibrium is induced due to the {\it surface drag} of the
exchange interaction. This is particularly relevant for
glass-like phases, where multiple spin configurations are
available. Indeed, it has been shown that magnetic frustration
and the existence of multiple easy axes orientations on the AFM
counterpart of the FM/AFM interface are essential for the
occurrence of training effects.\cite{hoffmann04}

The usual experimentally observed relationship between $M_E$ and
$\lambda$ is given by

\begin{equation}
M_E(\lambda) - M_E^{eq} \propto \frac{1}{\sqrt{\lambda}}
\label{Eq5}
\end{equation}
where $M_E^{eq}$ is the equilibrium magnetization shift. As shown
in the inset of Fig. 6, this equation holds for $\lambda \geq 2$,
with $M_E^{eq}=0.247$ Gauss. However, as already stated in
previous works\cite{binek04} this equation holds only for loop
indices $\lambda \geq 2$, and cannot explain the steep relaxation
between the first and second loops. In recent work
Binek,\cite{binek04} using a discretized relaxation model deduced
a recursive formula that relates the $(\lambda +1)$-th loop shift
with the $\lambda$-th one as

\begin{equation}
M_E(\lambda +1) = M_E(\lambda) -\gamma (M_E(\lambda) - M_E^{eq})^3
\label{Eq6}
\end{equation}
with $\gamma$ a sample dependent constant determined by
microscopic parameters like the spin damping constant. With this
simple relationship, and using $\gamma =17.5$ Gauss$^{-2}$ and
$M_E^{eq}=0.235$ Gauss, the whole set of data is reproduced (solid
triangles in Fig. 6), including the large initial drop of $M_E$.
From Eq. (\ref{Eq5}), the recursive law is recovered for $\lambda
\gg 1$, therefore explaining its applicability in this regime.

\section{Conclusions}

In summary, we provide the first evidence of intrinsic interface
exchange coupling in phase separated manganites. Our magnetization
measurements show the coexistence of a predominant charge-ordered
AFM phase and a minor FM component in the
Pr$_{1/3}$Ca$_{2/3}$MnO$_{3}$ compound. Furthermore, after cooling
the sample through the N\'{e}el temperature with an applied
magnetic field this manganite presents exchange bias. This is
clearly revealed by a distinctive magnetization shift ($M_E$) in
the remanence of the $M$-$H$ loops. The cooling field dependence
of $M_E$ at low temperature was successfully analyzed in terms of
a simple exchange interaction model between FM nanodroplets of
$\sim 10$ \AA$\,$in size and a disordered (glassy-like) AFM
shell. The observed training effect of the exchange bias can be
described with the same relaxation model used for other classical
exchange bias systems, indicating a similar mechanism for
exchange anisotropy. Moreover, the anisotropy constant deduced,
$K\sim 3\times 10^7$ erg/cm$^3$, is two orders of magnitude
larger than the magnetocrystalline anisotropy of other common FM
manganites. This large anisotropy energy opens a new possibility
for the development of hard FM materials, where important
exchange anisotropy interactions are introduced by intrinsic
inhomogeneities that naturally occur in phase separated magnetic
materials.

We acknowledge useful discussions with C. Leighton.

\end{document}